\documentclass[aps,prd,preprint,superscriptaddress,tightenlines,nofootinbib]{revtex4}

\usepackage{graphicx}
\usepackage{dcolumn}
\usepackage{bm}

\begin{document}

\preprint{CLNS 04/1893}       
\preprint{CLEO 04-13}         

\title{A New Measurement of the Masses and Widths of the $\Sigma_c^{*++}$ and $\Sigma_c^{*0}$ Charmed Baryons}


\author{S.~B.~Athar}
\author{P.~Avery}
\author{L.~Breva-Newell}
\author{R.~Patel}
\author{V.~Potlia}
\author{H.~Stoeck}
\author{J.~Yelton}
\affiliation{University of Florida, Gainesville, Florida 32611}
\author{P.~Rubin}
\affiliation{George Mason University, Fairfax, Virginia 22030}
\author{C.~Cawlfield}
\author{B.~I.~Eisenstein}
\author{G.~D.~Gollin}
\author{I.~Karliner}
\author{D.~Kim}
\author{N.~Lowrey}
\author{P.~Naik}
\author{C.~Sedlack}
\author{M.~Selen}
\author{J.~J.~Thaler}
\author{J.~Williams}
\author{J.~Wiss}
\affiliation{University of Illinois, Urbana-Champaign, Illinois 61801}
% ...
\collaboration{CLEO Collaboration} 
\noaffiliation


\date{October 27, 2004}

\begin{abstract} 

Using data recorded by the CLEO III detector at CESR, we 
have made measurements of some properties of the $\Sigma_c^{*++}$ and
$\Sigma_c^{*0}$ charmed baryons. 
In particular:
$\Gamma(\Sigma_c^{*++})=
14.4^{+1.6}_{-1.5}\pm1.4\ {\rm MeV},
\ M(\Sigma_c^{*++})-M(\Lambda_c^+) = 
231.5\pm0.4\pm0.3\ {\rm MeV},
\Gamma(\Sigma_c^{*0})=
16.6^{+1.9}_{-1.7}\pm1.4\ {\rm MeV},
\ M(\Sigma_c^{*0})-M(\Lambda_c^+) = 
231.4\pm0.5\pm0.3\ {\rm MeV}.$

\end{abstract}

\maketitle


In 1997\cite{SCS}, the CLEO collaboration reported the observation of two states, 
decaying into $\Lambda_c^+\pi^+$ and $\Lambda_c^+\pi^-$. 
These states were identified as the 
$\Sigma_c^{*++}$ and $\Sigma_c^{*0}$, the
cuu and cdd 
quark combinations in a $J^P = {3\over2}^+$ configuration.
Although the spin and parity of the states have not
been measured, it is generally assumed that this identification 
is correct.
In 2001\cite{SCSP}
CLEO showed the first evidence of the third member of the $\Sigma_c^*$ iso-triplet, 
the $\Sigma_c^{*+}$, which because of the
large background and low efficiency associated with the detection of a $\pi^0$, 
had a lower signal to noise ratio\footnote{
The $\Sigma_c^{*+}$ is not again addressed in this paper. The signal to background ratio
in the new data set is no better than in the old.}.
Although 
other experiments\cite{BELLE} have shown evidence for $\Sigma_c^*$ baryons, there have
been no further published measurements of their properties.
Measurement of the masses of the $\Sigma_c^{*++}$ and 
$\Sigma_c^{*0}$ can be used to check models that
predict the isospin mass splitting between these states, and measurements of their widths
can be compared with the widths of the $\Sigma_c$ baryons to check the predictions
of Heavy Quark Symmetry.

Here we present new measurements
of the natural widths of the $\Sigma_c^{*++}$ and $\Sigma_c^{*0}$ baryons, together with measurements
of their masses with respect to the $\Lambda_c^+$ mass. These measurements are made
using the CLEO III detector configuration\cite{RICH}. 
The CLEO III Ring Imaging CHerenkov (RICH)
detector enables a much superior separation of protons, kaon, and pions compared to previous versions 
of the CLEO detector. This in turn lowers the background for $\Lambda_c^+$ detection
and allows us to measure the properties of particles decaying into $\Lambda_c^+$
baryons with greater precision than was previously possible.

The data presented here 
were taken at the Cornell 
Electron Storage Ring,
corresponding to
an integrated luminosity of 14 $fb^{-1}$ 
in the energy range 9.4 to 11.5 GeV. 
This data set was acquired to study the decay products of the 
$\Upsilon(1S),\Upsilon(2S),\Upsilon(3S)$, and $\Upsilon(4S)$
resonances as well as the $e^+e^-\to q \bar{q}$ continuum data in this energy range.
The production mechanism of the charmed baryons is not under study here,
but we note that our kinematic cuts exclude the decay products of
$B$ mesons. Thus, the majority of the charmed baryons we measure 
are produced from the $e^+e^- \to q\bar{q}$ continuum, 
and some may be the direct decay products of the $\Upsilon$ resonances.

We observe the $\Sigma_c^*$ candidates by their decay 
$\Sigma_c^{*++/0} \to \Lambda_c^+\pi^{+/-}$.
Charge conjugate modes are implicit throughout.
For this measurement we use the two $\Lambda_c^+$
decay modes $\Lambda_c^+\to pK^-\pi^+$ and $\Lambda_c^+ \to p\overline{K^0}$. 
The CLEO III detector configuration detects charged particles
using a cylindrical drift chamber system inside
a solenoidal magnet.
Particle identification of $p,K$, and $\pi$ candidates is performed 
using specific ionization ($dE/dx$) measurements in the drift chamber, combined
with information, when present, from the RICH counters.
The technique for combining the two identification systems follows the method
that was used to find the decay $\Xi_c^0 \to pK^-K^-\pi^+$, and is described elsewhere\cite{XIC}.
The $\overline{K^0}$ candidates are found from the detection of the vertices
of $K^0_S\to \pi^+\pi^-$ decays that are significantly displaced from the beam spot.

To illustrate the good statistics and
signal to noise ratio of the $\Lambda_c^+$ signals, 
we reduce the combinatorial background, which is worse for
$\Lambda_c^+$ candidates with low momentum, by applying a cut on the
scaled momentum,
$x_p$. We define $x_p \equiv p_{baryon}/p_{max}$, where 
$p_{max} \equiv \sqrt{E^2_{beam}-M_{baryon}^2}$. 
We fit the invariant mass distributions for these modes to a sum
of a Gaussian signal and a low-order polynomial background. Figure 1 shows the 
plot for the $\Lambda_c^+\to pK^-\pi^+$ signal; the signal yield is approximately 
45,000. 
The decay mode $\Lambda_c^+ \to p\overline{K^0}$, which has superior signal to 
noise ratio, augments the number of candidates by 15\%.

Releasing the $x_p(\Lambda_c^+)$ cut, $\Lambda_c^+$ candidates within $2.0\ \sigma$ of the 
peak mass in each decay mode were combined with each remaining
charged $\pi$ track in the event. 
A cut of $x_p > 0.5$ was made on the $\Lambda_c^+\pi$ combination, 
and the mass difference $\Delta(M)=M(\Lambda_c^+\pi)-M(\Lambda_c^+)$ was calculated. 

The mass difference spectra, shown in Figure 2, are plotted in the 
mass range 178-298 MeV.
The lower bound of this plot is chosen to avoid the contribution of
$\Sigma_c^{++/0} \to \Lambda_c^+\pi^{+/-}$ decays. The upper bound is chosen 
to approximately center the $\Sigma_c^*$ peaks.

The fits to the signal spectra in Figure 2 each have three components as follows.
Firstly, the excesses in the 
region below 204 $\rm{MeV}$ due to $\Lambda_{c1}^{+}(2625)$ production are
accommodated by 
functions found using the $\Lambda_c^+\pi$ spectrum from 
reconstructed $\Lambda_{c1}^{+}(2625)$ decays, with the normalization 
corrected using a Monte Carlo program. 
Secondly, we use a background shape of a first order polynomial.
Lastly,  we use signal functions 
of spin-1 Breit-Wigners convolved with a Gaussian resolution function 
of standard deviation, $\sigma=1.61\ \rm{MeV}$. 
This resolution was calculated using a GEANT-based 
Monte Carlo simulation program\cite{GEANT} for a peak in the region under scrutiny.

In the case of $\Lambda_c^+\pi^+$ we obtain a signal 
of $1330\pm110$ events, with a width of 
$\Gamma=14.4^{+1.6}_{-1.5}\ \rm{MeV}$, and a mass difference of
$\Delta(M)=231.5\pm0.4\ \rm{MeV}$.
For the $\Lambda_c^+\pi^-$, we obtain a signal 
of $1350\pm120$ events, with a width of 
$\Gamma=16.6^{+1.9}_{-1.7}\ \rm{MeV}$, and a mass difference of
$\Delta(M)=231.4\pm0.5\ {\rm MeV}$.

We have checked that consistent widths and mass differences are extracted 
if the data are restricted to particles or anti-particles, and if the data
are restricted to data taken at the $\Upsilon(1S),\Upsilon(2S)$ and $\Upsilon(3S)$ 
(resonances where the signal to background ratio is not as high), or in the $\Upsilon(4S)$ and
non-resonant regions.

The parameters of the 
extracted signals depend on the exact method of fitting used. 
We have tried many variations of the background functions and fitting range.
These include using a second-order Chebychev polynomial for the background function,
allowing the $\Lambda_{c1}(2625)$ contribution to float, and extending the fitting 
range to much higher values of $\Delta(M)$.
The systematic uncertainties
in the measurements due to the fitting procedures 
are taken as being the maximum range of parameters 
obtained using different reasonable fits of these types. This is the dominating 
systematic uncertainty for the width measurements.
The results of the Monte Carlo simulation program that predicts the detector resolution
have been checked using a series of 
narrow states in our data. Based on the agreement of the simulation and the results
of analyzing these peaks, we assign a systematic uncertainty of $\pm 0.13 \ $ MeV
to the value of the detector resolution for the $\Sigma_c^*$ states. 
As the resolution is an order of magnitude 
less than the intrinsic widths that we measure, this uncertainty produces a modest
uncertainty ($\approx 0.14\ $ MeV) in the measurements of the widths, which is 
negligible compared with the uncertainty due to the fitting technique.

The masses of the signals are insensitive to the fitting technique, 
and the different fits produce a maximum range 
of $\pm 0.2\ $ MeV for the measurements. 
We allow for a systematic uncertainty of 0.2 MeV from possible analysis biasses.
An uncertainty in the CLEO magnetic field strength of 0.1\%, which is bounded by 
measurements of particles of known mass, corresponds to an uncertainty of 0.13 MeV
in the mass difference measurements. These three sources of uncertainty 
add in quadrature to give a total systematic uncertainty in the mass difference 
measurements of 0.3 MeV. These last two uncertainties do not contribute
to the systematic uncertainty in the isospin mass splitting between the two states.

Our measurements of the $\Sigma_c^*$ masses and widths are consistent with
the previously published CLEO numbers\cite{SCS}. However, the splitting between 
the two states $M(\Sigma_c^{*++})-M(\Sigma_c^{*0})$, is now measured to be 
$-0.1\pm0.8\pm0.3\ $MeV, whereas previously it was $+1.9\pm2.0$ MeV. This isospin 
splitting is expected to be small and negative\cite{PRED}. 
The difference in the widths of the two states is expected to be negligible. 
By Heavy Quark Symmetry\cite{PIRJOL} 
the ratio {$\Gamma(\Sigma_c^*)/{ \Gamma(\Sigma_c)}$} should 
equal
$({M_{\Sigma_c}/{M_{\Sigma_c^*}}}) \times ({p_{\pi}(\Sigma_c^*)^3/{p_{\pi}(\Sigma_c)^3}})$,
where $p_{\pi}$ is the momentum of the $\pi$ in the parent's rest frame. The isospin mass
splitting of the states is very small, and for both the doubly charged and neutral states
this latter quantity equals $7.5\pm0.1$.  Using our new values of the $\Sigma_c^{*}$ widths,
and world average values of the $\Sigma_c$ states\cite{PDG}, we obtain
${\Gamma(\Sigma_c^{*++})/{ \Gamma(\Sigma_c^{++})}} = 6.5\pm1.3$ and
${\Gamma(\Sigma_c^{*0})/{ \Gamma(\Sigma_c^{0})}} = 7.5\pm1.7$, in excellent
agreement with expectation.

In conclusion, we present new measurements of the properties of the
$\Sigma_c^{*++}$ and $\Sigma_c^{*0}$ baryons. 
For the doubly charged state
$M(\Sigma_c^{*})-M(\Lambda_c^+)$ is measured to be $231.5\pm0.4\pm0.3\ {\rm MeV}$ 
and $\Gamma=14.4^{+1.6}_{-1.5}\pm1.4\ {\rm MeV}$, and for the neutral state
$M(\Sigma_c^{*})-M(\Lambda_c^+)$ is measured to be $231.4\pm0.5\pm0.3\ {\rm MeV}$ 
and $\Gamma=16.6^{+1.9}_{-1.7}\pm1.4\ {\rm MeV}$.
\smallskip
\centerline{\bf ACKNOWLEDGMENTS}
\smallskip

\begin{figure}
\includegraphics*[width=3.75in]{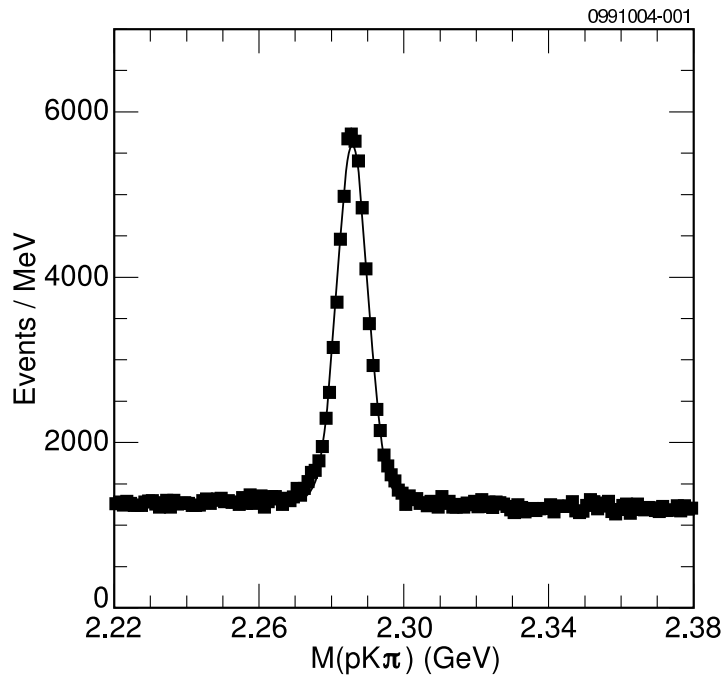}
\caption{Invariant mass plot for $pK^-\pi^+$ combinations with $x_p(pK^-\pi^+)>0.45$}
\label{fig:pkpi}
\end{figure}
\begin{figure}
\includegraphics*[width=4.25in]{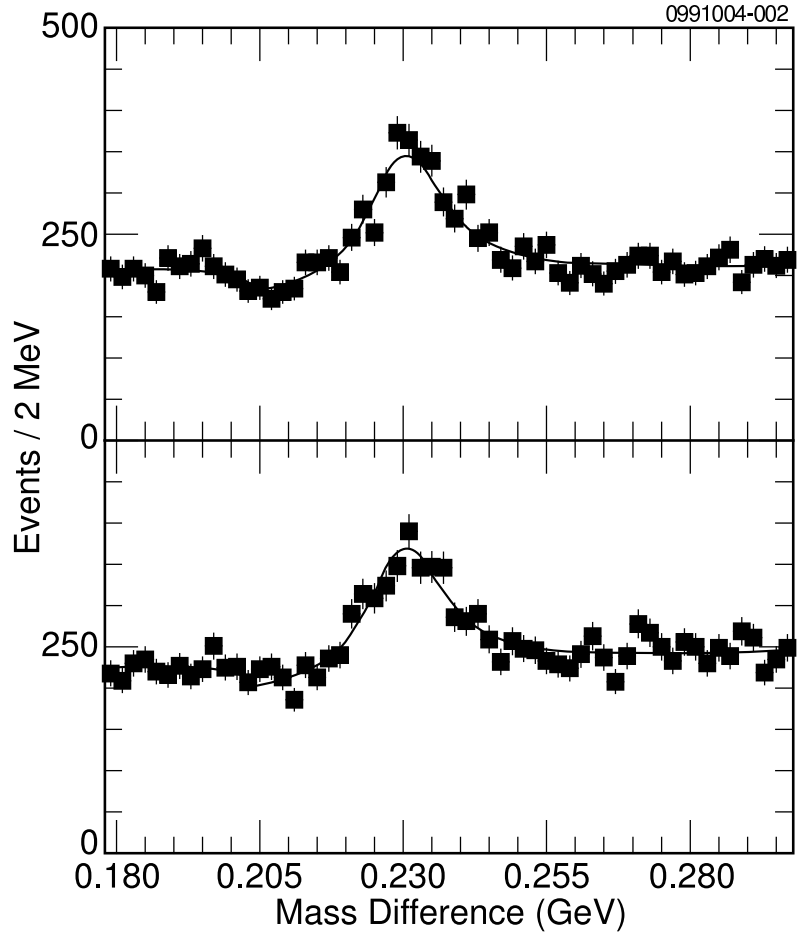}
\caption{$M(\Lambda_c^+ \pi)-M(\Lambda_c^+)$ mass differences for (upper) $\pi^+$
and (lower) $\pi^-$ combinations}
\label{fig:sigcs}
\end{figure}

We gratefully acknowledge the effort of the CESR staff 
in providing us with excellent luminosity and running conditions.
This work was supported by the National Science Foundation
and the U.S. Department of Energy.

\end{document}